\documentstyle[prb,aps,twocolumn]{revtex} 
\input psfig
\begin{document}

\newcommand{\be}{\begin{equation}}
\newcommand{\ee}{\end{equation}}
\newcommand{\bea}{\begin{eqnarray}}
\newcommand{\eea}{\end{eqnarray}}
\newcommand{\dif}{{\mathrm d}}

\author{N. I. Lundin$^{1,2}$ and Y. M. Galperin$^{1,3}$}
\title{Impurity-induced dephasing of Andreev states}
\address{$^1$Centre for Advanced Study, Drammensveien 78, 0271 Oslo, Norway.\\
$^2$Department of Applied Physics, Chalmers University of
Technology and G\"oteborg University, 412 96 G\"oteborg, Sweden\\
$^3$Department of Physics, University of Oslo, P. O. Box
1048, N-0316 Oslo, Norway and\\
Division of Solid State Physics, Ioffe
Institute of the Russian Academy of Sciences,
St. Petersburg 194021, Russia
\date{\today}
{\parbox{14cm}
{\small
\vspace*{0.5cm}
A study is presented concerning the influence of flicker noise in the
junction transparency on coherent transport in Andreev states.  
The amount of dephasing is estimated for a microwave-activated quantum
interferometer. Possibilities of experimentally investigating the
coupling between a 
superconducting quantum point contact and its electromagnetic
environment are discussed.}}}

\maketitle

\section{Introduction}

How much can flicker noise in the junction transparency affect coherent
transport in Andreev states present in a superconducting quantum point
contact (SQPC)? The assumption of coherent transport in Andreev states 
is widely used in theoretical work, see e.~g. Refs.~\onlinecite{averin}
and~\onlinecite{bratus}. However, in
realistic systems interaction with a dynamical environment will always
introduce some amount of dephasing, see 
Refs.~\onlinecite{imry} and~\onlinecite{ingnaz} for a review. 

In the so-called 
microwave-activated quantum interferometer~\cite{prl} full coherency of 
Andreev states is assumed and a method for Andreev-level spectroscopy
is presented. The Andreev levels are probed with a microwave field, resulting
in an interference pattern in the current. If dephasing is present 
this interference pattern will begin to deteriorate. This connection 
between dephasing and current makes the microwave-activated quantum 
interferometer a suitable system to study the effect of flicker noise in
the junction transparency on transport through Andreev states.
It also provides an excellent opportunity to probe the coupling
between a SQPC and its electromagnetic environment. 

In the following we study the role of  dephasing induced by flicker
noise in the normal state junction transparency,  
$D$, of a SQPC in the transport through Andreev states. The concrete
system which will be considered is the above presented
interferometer.~\cite{prl}   
Flicker noise can be caused by the presence of an impurity atom 
close to the junction which has two states of almost 
equal energy to choose from. When the atom tunnels between its two
states the junction transparency will fluctuate. Another source of
flicker noise is the tunneling of an  
electron between impurity atoms in a doped region. If there are two
such neighboring defects with available states, a hybrid two-level
state is formed and the electron can hop between the two. This hopping
will then add a fluctuation to the junction transparency. The
amplitude of these fluctuations depend on the distance between the
defects and the junction. From now on we will refer to these
dynamic defects as two-level elementary fluctuators (EFs).

\begin{figure}[h]
\centerline{\psfig{figure=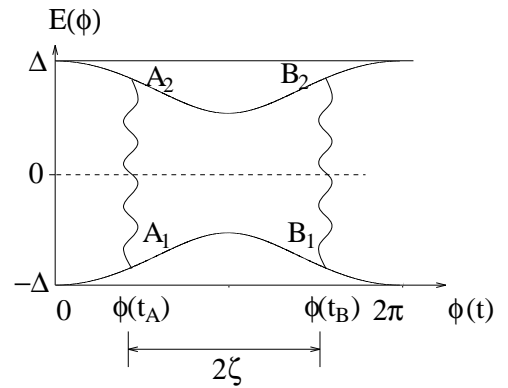,width=6cm}}   
\vspace*{0.2cm}
\caption{Time evolution of the Andreev levels within the energy gap of
a single-mode SQPC. The wavy lines connecting points $A_1$, $A_2$ and
$B_1$, $B_2$ symbolize resonant transitions between the levels induced
by an applied microwave field. The symbol $\zeta$ defines the position
of the resonances. \label{fig:spectra}}
\end{figure}

The microwave-activated quantum interferometer (MAQI) is based on a
short, weakly biased SQPC which is subject to microwave
irradiation. Confined to the contact area  
there are current carrying Andreev states. The corresponding 
energy levels -- Andreev levels -- are found in pairs within the
superconductor  
energy gap $\Delta$, 
one below and one above the Fermi level. When a SQPC is short
($L\ll\xi_0$ where $L$ is the length of the junction while $\xi_0$ is
the superconductor coherence length), there is only one pair of
Andreev levels and their positions depend on the
order parameter phase difference, $\phi$, across the contact as
\be
E_\pm=\pm E(\phi)=\pm\Delta\sqrt{1-D \sin^2(\phi/2)}.
\label{eq:spectra}
\ee
Within this pair,
the two states carry current in opposite directions and in equilibrium only
the lower state is populated. The applied bias, $V$, through
the Josephson relation $\dot{\phi}=2eV/\hbar$,
forces the Andreev levels to move adiabatically within the energy gap
with a period of $T_p=\hbar\pi/eV$, see Fig.~\ref{fig:spectra}.

The microwave field induces Landau-Zener (LZ) transitions between 
the Andreev levels (symbolized by wavy lines in Fig.~\ref{fig:spectra}).
If the upper level is populated after the second transition a
delocalized quasiparticle excitation will be created when this Andreev
level merges with the continuum. The result will be a dc contribution to the
current. Further, this current exhibits an interference pattern  since there
are two paths available to the upper level. It is this ``interference
effect'' which is utilized in the MAQI for Andreev-level spectroscopy.

In order to estimate the influence of flicker noise on the
interference pattern we continue with a quantitative presentation of the
interferometer and the model used for the fluctuations.

\section{Model and method}
Consider a short single-mode SQPC which is subject to a high frequency
microwave field ($\hbar\omega=2E(\phi)\le2\Delta$). Let the contact,
placed at $x=0$, be 
characterized by an energy-independent transparency $D$. A weak bias,
$eV\ll\Delta$, is applied across the junction. We choose to describe the
quasiparticles in the contact region with the following wave function,
\be
\Psi(x,t)=u_+(x,t)e^{ik_Fx}+u_-(x,t)e^{-ik_Fx},
\ee
where the envelope functions $u_\pm(x,t)$, left and right movers, 
are two-component vectors in electron-hole space. To simplify notation we
introduce the four-component vector ${\bf u}=[u_+, u_-]$. This vector
satisfies the
time-dependent Bogoliubov-de Gennes  equation 
\begin{eqnarray}
i\hbar \partial{\bf u}/\partial t &=&[{\cal H}_0  + \sigma_z\, V_g(t)]
{\bf u} \,, 
\label{BdG} \\
{\cal H}_0&=&-i\hbar v_F\sigma_z\tau_z\, \partial / \partial x 
+ \Delta\left[\sigma_x \cos (\phi(t)/ 2)
\right. \nonumber \\  && \left. 
 + \,  \mbox{sgn} \,(x) \, \sigma_y\, 
\sin(\phi(t) / 2)\,\right]\, , 
\label{H0}
\end{eqnarray}
where $\sigma_i$ and $\tau_i$ denote Pauli matrices in
electron-hole space and in $\pm$ space, respectively, while $
V_g(t)=V_\omega\cos(\omega 
t)$ is the time-dependent gate potential. We
assume $eV_\omega \ll \Delta$.

The boundary condition at $x=0$ is 
$${\bf u}(+0)=D^{-1/2}\left[1-\tau_y\, (1-D)^{1/2}\right]\,{\bf
u}(-0)\, .$$
As a result of the applied high frequency field there will be 
resonant transitions between the Andreev levels. These transitions
introduce a mixed state which can be described within the resonance
approximation as
$${\bf u}(x,t)=\sum_\pm b^\pm (t)\, {\bf u}^\pm (x) \,e^{\mp \omega
t/2}\, , $$ 
where ${\bf u}^+(x)$ and ${\bf u}^-(x)$ are the envelope functions of
the upper 
and lower Andreev states, while
$b^+$ and $b^-$ are the corresponding probability amplitudes.  The
final result is a dc current through the SQPC,\cite{prl}  
\begin{eqnarray}
I_{\rm
dc}&=&\frac{e(2\Delta-\hbar\omega)}{\hbar\pi}
\left|b^+\left(
{\hbar\pi \over eV}\right)\right|^2   
=2I_0\sin^2{(\Theta+\Phi)}, \nonumber \\
I_0 &\equiv& (2 e/\hbar\pi)\,  r^2(1 -
 r^2 )\left(2\Delta-\hbar  \omega \right)\, ,
\label{curr}
\end{eqnarray}
where $r^2$ is the LZ transition amplitude, which depends on
the bias voltage and the amplitude of the perturbation ($\Theta$ is
the phase of the LZ transition, which can be considered constant).\cite{prl}
The phase $\Phi$ which inhibits the interference is calculated through 
 \be
  \Phi=\frac{1}{2eV}\int_{\phi_A}^{\phi_B}\left( E (\phi)-
 \frac{\hbar\omega}{2}\right)\dif\phi\, . 
 \label{int_phase}
 \ee
Flicker noise of the junction transparency, $D$, which is the subject to 
be discussed now, enters through $E(\phi)$, Eq.~(\ref{eq:spectra}), in
this expression. 

\subsection{Flicker noise in the junction transparency}
The main topic of this work is to study the effect of fluctuations
in the junction transparency on the MAQI. For simplicity we choose
to model the sources of these fluctuations, the EFs, with the
so-called {\emph {random telegraph process}}. This process is
characterized by a random quantity $\xi(t)$ which has the value
$+1$ or $-1$ depending on whether the upper or lower EF state is 
occupied. We assume that the probability of each state is the same,
namely $1/2$. 
This is acceptable since EFs with inter-level distances,
$E_i$, smaller than $k_{\text{B}} T$, will be ``frozen'' --- they
behave as static impurities
which do not affect the dynamic fluctuations of $D$.
In this model the EF
switches between its two states randomly in time. Physically,
switching is a result of interactions between the EF and phonons or
electrons in the contact area.  

In the presence of EFs the junction transparency will be
modulated. In other words, $D\rightarrow D+D_f(t)$, where $D_f(t)$ is
assumed to be small. Generally 
$$D_f(t)=\sum_i A_i \, \xi_i(t)\, , $$ with
$A_i$ being the coupling strength of the $i$-th EF. We assume that the
random processes in different EFs are not correlated. Consequently,
after a change in 
variables from $t$ to $\phi$ we can specify the random telegraph
processes $\xi_i(t)$ through the correlation function,
\be
\langle \xi_i(\phi_1)\xi_j(\phi_2)\rangle=\delta_{ij}
e^{-2\gamma_i|\phi_2-\phi_1|},
\ee
where $\gamma_i$ is the switching rate of $i$-th EF in units of
the Josephson frequency $\omega_J=2eV/\hbar$. It is related to the
dimensional switching rate $\Gamma$ as $\gamma \equiv \hbar \Gamma /2eV$. 

Depending on the construction of the SQPC there can be any number of
EFs which are ``in range'' to influence the transparency. In 
junctions which are very small it is probable that only one single
EF will be in the vicinity of the contact. In this case, the
coupling constant $A$ and the switching rate $\Gamma$ can be directly
evaluated from the measured telegraph noise intensity in the normal state,
$$S(\tau) \equiv \langle I(t+\tau)I(t)\rangle_t- \langle
I(t)\rangle^2_t\, . $$
Indeed, the current through a single mode QPC at low
temperatures can be expressed according to the Landauer formula as  $I(t)=2 e^2
V D(t)/h$. Consequently, the random telegraph noise intensity 
is equal to 
\be
S(\tau)=(2 e^2V A/h)^2\, \exp (-2 \Gamma \tau)\, ,
\ee  
and both $A$ and $\Gamma$ can be extracted from measured $S(\tau)$. A
possible approach for extracting these model parameters from noise
measurements in the case of many fluctutators in the QPC area will be
discussed later.
 
\section{Single EF}\label{single}

In the case of a very small contact it is possible to consider only
one EF and put $i=1$.  We start by decomposing Eq.~(\ref{eq:spectra}) as
\bea
&&E(\phi)=E_+(\phi)+E_-(\phi)\xi(\phi),\nonumber \\
&&E_\pm \equiv \frac{1}{2}\left[E(\phi|D_1)\pm
E(\phi|D_{-1})\right], \label{decomposed} 
\eea 
where $D_{\pm1}$ are the two different values the transparency 
fluctuates between. Further, we assume that both $\gamma$ and 
$g_{fs} \equiv E_-/eV$ are much smaller than
the reduced inter-level distance $(E_+\approx\Delta)/eV$. This means that all
deviations in time are much longer than the Andreev level 
formation time, which is of the order $\hbar/2\Delta$.

Fortunately, the expression above, Eq.~(\ref{decomposed}), is linear
in $\xi$ and we can write the effect of the EF as an additive
contribution to the accumulated phase, Eq.~(\ref{int_phase}), without
making any approximations. Namely, $\Phi\rightarrow\Phi+\Phi_f$, with
\be
\Phi_f=\frac{1}{2}\int_{\phi_A}^{\phi_B} g_{fs}(\phi)\, \xi(\phi)\, \dif\phi. 
\ee
with $g_{fs} (\phi)=E_-(\phi)/eV$.
After averaging over the realizations of the random process $\xi (t)$,
the expression ~(\ref{curr}) for the MAQI current is replaced by 
\be
I_{\rm dc}=I_0\, [1-W \cos{(2\Theta+2\Phi)}], 
\label{curr_fl}
\ee
where $W \equiv\langle e^{2i\Phi_f} \rangle$ contains the dephasing.
Without dephasing $W=1$ and the effect of the phase $\Phi
(\phi_A,\phi_B)$ is a 
modulation of the dc current between 0 and $2I_0$, the modulation
depth
\be
(I_{\max} - I_{\min})/(I_{\max} + I_{\min})=|W| \, . 
\label{md}
\ee
being equal to one.
 This modulation of the current is the interference effect utilized in
the MAQI. When dephasing  enters, the quantity $W(\phi_A,\phi_B)$ will
decrease and the modulation $|W|$ 
of the current envelope, i. e. the interference pattern,  will in turn
decrease. 

To facilitate the calculation of the dephasing term, $W$, we define the
auxiliary function,  
\be
\Psi(\phi)=\left< \exp\left[i\int_{\phi_A}^\phi \dif \phi' \, 
g_{fs}(\phi')\, \xi(\phi')\right]\right>\, . \label{aux_func}
\ee 
 The quantity of interest, $W$, is related to $\Psi(\phi)$ as 
$
W\equiv \Psi(\phi_B)$.
Further, it can be shown, cf. with  Ref.~\onlinecite{brissaud},
that Eq.~(\ref{aux_func}) satisfies the
differential equation
\be
\frac{\dif^2\Psi}{\dif \phi^2}+\left[ 2\gamma- \frac{\dif\ln
    g_{fs}(\phi)} {\dif \phi}\right]\frac{\dif \Psi}{\dif \phi} +
g_{fs}^2(\phi)\Psi=0,\label{diff_eq}
\ee
with the initial conditions
\be
\Psi(\phi_A)=1\, , \quad \dif\Psi/\dif\phi|_{\phi=\phi_A}=0\, . \label{ic}
\ee 
Let us consider the following two cases in more detail:
(i) the ``slow EF'',
$\gamma \ll g_{fs}$, which corresponds to low temperatures, and (ii)
 the ``fast EF'', $\gamma \gg g_{fs}$, which corresponds to relatively
high temperatures.

\subsection{Slow switching EF}\label{single_slow}
In the low-temperature limit the EF will slowly switch between its two
states, and we can let $\gamma\rightarrow 0$ in
Eq.~(\ref{diff_eq}). Further, by introducing the function,
\be
\Xi(\phi)=\int_{\phi_A}^{\phi}\dif \phi \, g_{fs} (\phi)\Psi(\phi),
\label{xi_func}
\ee
and applying the initial conditions (\ref{ic}) we obtain the integral equation,
$$\Xi^2+\Psi^2=1 \, .$$
The solution follows as 
$\Psi^{(0)}(\phi)=\cos \Phi(\phi)$ where $\Phi(\phi) \equiv \left(\int_{\phi_A}^\phi \! \dif \phi'\, g_{fs} (\phi')
\right)$. Consequently,  $W^{(0)}=\cos \Phi_\zeta$ with
\be
\Phi_\zeta \equiv \Phi(\phi_B)=\int_{\phi_A}^{\phi_B}\! \dif \phi \,
g_{fs} (\phi)\, . 
\label{eq:Phi_f}
\ee
A better approximation can be found by looking for the solution in the
form $\Psi^{(1)}(\phi)=u(\phi)\Psi^{(0)}(\phi)$ and assuming $u(\phi)$
to be a slow function. Neglecting the second derivative of $u(\phi)$ we
obtain the differential equation 
\be
\dif u/\dif \phi =- \gamma \,
\eta(\phi)\, u
\label{de1}
\ee
  for $u(\phi)$ with
\be 
\eta(\phi) \equiv \left(1+\frac{1}{2g_{fs}^2\tan \Phi(\phi)}\frac{\dif
g_{fs}}{\dif \phi}\right)^{-1}\, . \label{eta_phi}
\ee
Defining 
\be \label{eq:Uc}
\upsilon_\zeta \equiv \frac{1}{2\zeta}\int_{\phi_A}^{\phi_B}\eta(\phi)\, \dif
\phi
\ee
we arrive at the following 
expression for the oscillating part of the current,
\bea
&&W\cos (\Phi)=e^{-2\gamma \upsilon_\zeta \zeta} \cos (\Phi_\zeta) \cos(\Phi) 
\nonumber \\ && = 
e^{-2\gamma \upsilon_\zeta \zeta}\, \left[\cos(\Phi+\Phi_\zeta)+
\cos(\Phi-\Phi_\zeta)\right]/2 \,. \label{eq:W_slow}
\eea
For a constant $g_{fs}$, $\upsilon_\zeta=1$ and $\Phi_\zeta=2g_{fs}\zeta$,
where $\zeta=(\phi_B-\phi_A)/2$ is equal to half the distance between
the resonance positions, see Fig.~\ref{fig:spectra}. In the general
case these quantities are increasing functions of $\zeta$.
 
\begin{figure}[h]
\centerline{\psfig{figure=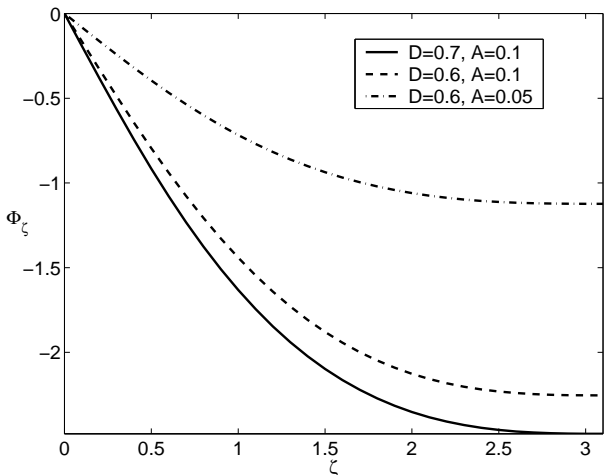,width=8cm}} 
\caption{In the case of flicker noise from a single EF in the low
temperature limit the MAQI interference pattern is split into two
parts, phase-shifted by $\Phi_\zeta$, shown here as a function of the
resonance position $\zeta$. The inset shows the contact transparency
and the strength of the fluctuator. Here $eV/\Delta\approx 0.1$. \label{fig:single_slow}}
\end{figure}

At $\gamma=0$, the current is split into two interference patterns of
equal magnitude 
shifted by the phase $\Phi_\zeta$. A plot of $\Phi_\zeta$ as a function of $\zeta$ is presented in Fig.~\ref{fig:single_slow}.
The transparency $D$ and the strength of the fluctuator are shown in the
inset. There is no dephasing, only a distortion of the interference pattern, 
the modulation being $|W|=|\cos \Phi_\zeta|$.

\begin{figure}[h]
\centerline{\psfig{figure=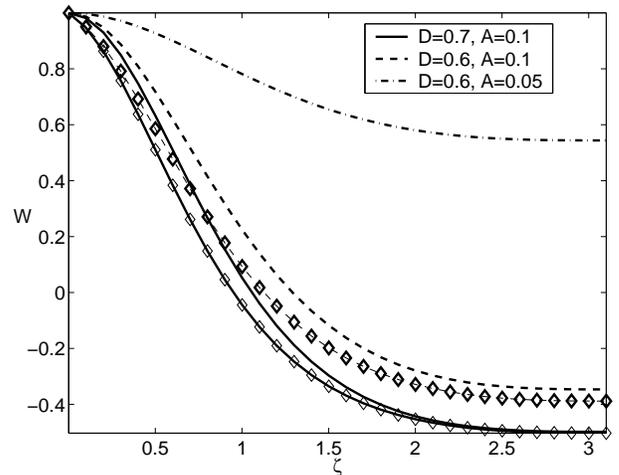,width=8cm}} 
\caption{In the case of flicker noise from a single EF at low
temperatures, when the hopping rate $\gamma$ is non-zero, there will
always be some amount of dephasing (see text). This figure shows the
dephasing factor $W$ 
as a function of the resonance position
$\zeta$ when $\gamma=0.1$. The two curves marked with diamonds ($\diamond$) are
calculated from $W$ in Eq.~(\ref{eq:W_slow}), while the rest are direct
numerical solutions of Eq.~(\ref{diff_eq}).
 The inset shows the contact transparency
and the strength of the fluctuator. In this plot $eV/\Delta\approx 0.1$. \label{fig:single_M}}
\end{figure}

This splitting into two patterns of equal magnitude follows from the 
assumption that the occupation probability is the same 
for the two EF states.
The general case of arbitrary probabilities for the EF states
can be solved numerically. At finite $\gamma$ dephasing takes place
and the amplitude of the interference oscillations decreases by $e^{-2
\gamma \upsilon_\zeta \zeta}$. The physical reason of dephasing is the
finite life 
time of an EF in a given state. We have calculated $W$ for
$\gamma=0.1$, comparing the analytical approximation above against a
numerical solution of Eq.~(\ref{diff_eq}), cf. with Fig.~\ref{fig:single_M}. 
(The plot for the analytical case when $D=0.6, A=0.05$ is missing because of
numerical difficulties).
%

\subsection{Fast switching EF}\label{single_fast}
In the case of fast switching the differential equation (\ref{diff_eq})
can be approximated as,
\be
\frac{\dif \Psi}{\dif \phi} =-\frac{g_{fs}^2(\phi)}{2\gamma}\Psi(\phi).
\label{diff_eq_fast}
\ee
The solution in this case is easily obtained, as 
\be
W=\Psi(\phi_B)=\exp[-K(\phi_B,\phi_A)],
\label{eq:W_fast}
\ee
with
\be
K(\phi_B,\phi_A)=\frac{1}{2\gamma}\int_{\phi_A}^{\phi_B}\dif \phi\,
g^2_{fs}(\phi). \label{eq:k_phi}
\ee 
\begin{figure}[h]
\centerline{\psfig{figure=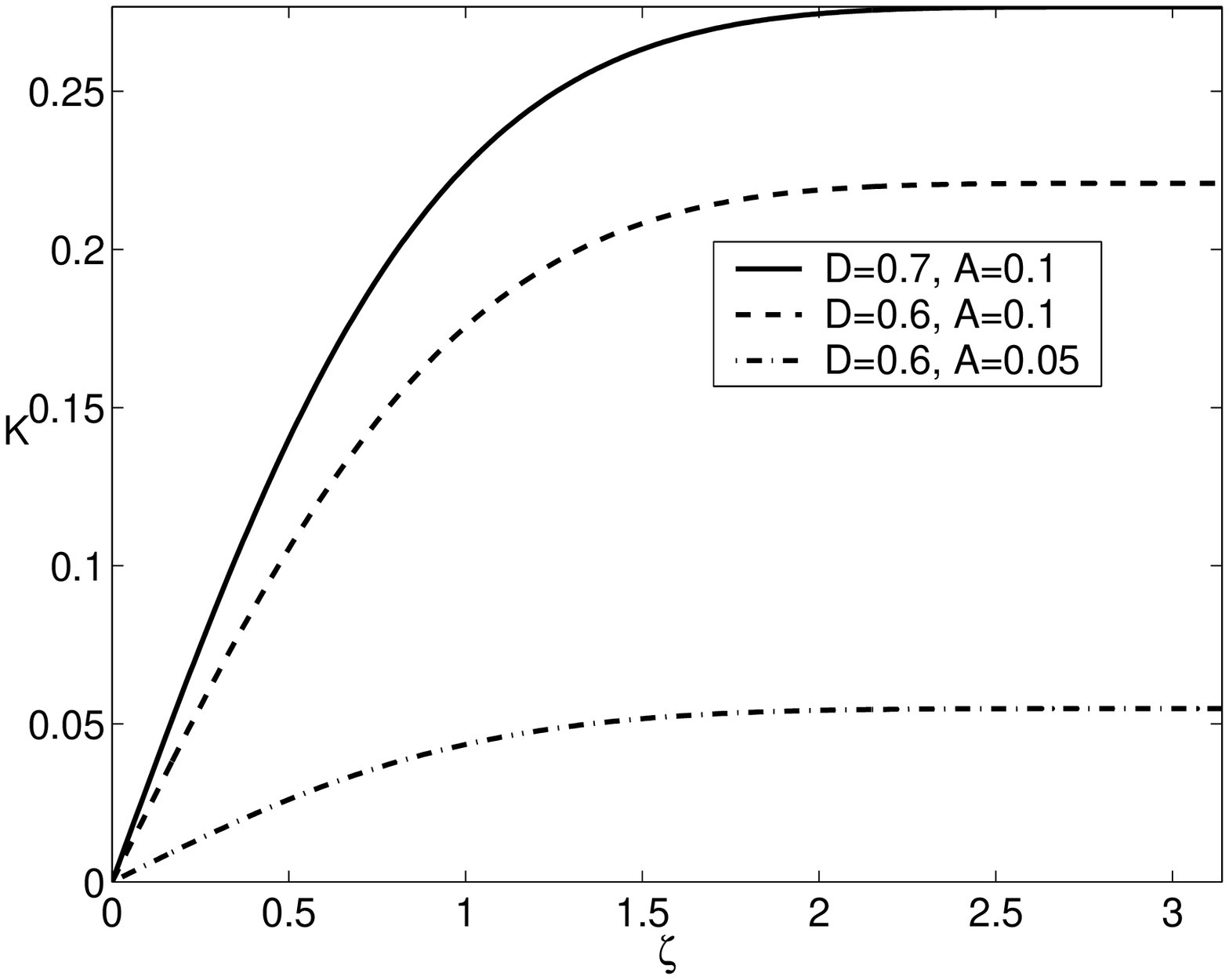,width=8cm}}
\caption{In the high temperature limit, flicker noise of a single EF
leads to damping of the MAQI interference pattern by a factor of
$W=\exp(-K)$. The inset shows the contact transparency
and the strength of the fluctuator. In this plot $eV/\Delta\approx
0.1$ and the switching rate is $\Gamma=3$.} \label{fig:single_fast}
\end{figure}
Here we find an exponential decay of the interference term. 
Expression (\ref{eq:k_phi}) describes an  effect which is similar to {\em
motional narrowing} of spectral lines.\cite{klauder}
When the EF fluctuates rapidly enough comparied to the ``energy
resolution'' $E_-/\hbar$, influence from the difference
between two EF states is smeared and dephasing will be of a diffusive
character, with en effective, time-dependent, diffusion constant $g_{fs}^2/2\gamma$. A calculation of the factor $K$ for $\gamma=3$ 
is shown in Fig.~\ref{fig:single_fast} for three different 
sets of values of the junction transparency and the EF's strength. 
The effect of dephasing from flicker noise will decrease for 
higher bias voltages, since a higher bias, through the 
Josephson relation, leads to a shorter time between the resonances.

To get an idea about the general case of arbitrary switching rates, it
is instructive to consider a $\phi$-independent $g_{fs}$.  This approximation is only
valid when $|\dif E(\phi)/\dif\phi|$ is small, but should provide the
general behavior of $W$.  In this limit Eq.~(\ref{diff_eq}) can be
solved analytically, resulting in
\bea \label{eq:W_single}
&&W=e^{-2\gamma\zeta}\left[
\cosh\left(2\zeta\sqrt{\gamma^2-g_{fs}^2}\right)+\right.\nonumber\\ 
 &&\left.\frac{\gamma}{\sqrt{\gamma^2-g_{fs}^2}}\sinh\left(2\zeta
\sqrt{\gamma^2-g_{fs}^2}\right)\right]\,  .
\eea
From this expression one can conclude that there will always enter an
exponential decay of the interference current, except for the limit of
$\gamma\rightarrow 0$. In this limit there should only
be an additional oscillation added to the current as a function of
$\zeta$ since for
$\gamma \le |g_{fs}|$, the hyperbolic functions
behave as trigonometric functions.

\section{Many EF's}

Let us consider a large number of EFs with varying switching rates
distributed in the contact area. 
For simplicity, we shall assume that only the fluctuators with
inter-level spacings $U_i \lesssim kT$ are important, and that their
distribution is uniform, ${\cal P}_U (U)=P_0 {\cal V}$. Here $\cal V$
is the sample volume. Further, we
assume that the switching rates $\gamma_i$ are the same for
both transition directions (up och down) between the EF's levels. This
assumption is natural because the ratio between the corresponding
transition rates is $\exp(-U_i/kT)$. Within the assumptions discussed
above, the final results are substantially simplified while preserving
the essential depedence on temperature and the resonance
position. These approximations agree with a general theory 
developed in  Ref.~\onlinecite{laikhtman} for the case of dephasing by
two-level systems (TLS) in glasses. 

The first step now is to 
linearize the SQPC's transparency with respect to $\xi_i$ as $D\rightarrow
D+D_f(t)$.  This allows
us to once again find an additive contribution to the accumulated
phase, Eq.~(\ref{int_phase}), which in this case will be,
\be
\Phi_f\approx\sum_iA_i\int_{\phi_A}^{\phi_B}\!
\xi_i(\phi)\, g_{fm} (\phi)\, \dif\phi\, .
\ee
Here we have defined $g_{fm}(\phi)=(1/2eV)\dif E(\phi)/\dif D$.
In the same manner as in Sect.~\ref{single}, but this time averaging
over $A$'s, $\gamma$'s and $\xi's$, we can express the modified MAQI
current through  expression~(\ref{curr_fl}) with 
\be 
W=\left\langle e^{i\sum_{i}A_i \int_{\phi_A}^{\phi_B}\!
g_{fm}(\phi)\, \xi_i(\phi)\, \dif\phi}
\right\rangle_{\! A,\gamma,\xi}  \, .
\label{eq:W}
\ee
To approximate this average we use the Holtsmark method~\cite{Holt} which is
valid in the limit of many fluctuators, $N=P_0\,{\cal V}\, kT \gg
1$. This allows us to rewrite 
Eq.~(\ref{eq:W}) as the average over the contributions from single EFs,
\be
W_{s}(A, \gamma)= \left\langle \exp\left(i A \int_{\phi_A}^{\phi_B}\!
g_{fm}(\phi)\, \xi(\phi)\, \dif\phi \right)\right\rangle_{\xi}
\ee
as
\be
W\approx\exp\left(-P_0{\cal V} kT\left\langle 1-
W_{s}(A,\gamma)\right\rangle_{A,\gamma}\right)\, .\label{eq:W_h} 
\ee
Since the number of EFs is assumed to be large, to keep dephasing at
a reasonable level it is important to keep $\langle1-W_s\rangle$
small.

With the solutions for $W_{s}$ found in Sect.~\ref{single} the
average $\langle 1-W_{s}\rangle$ remains to be calculated. To
calculate this average one has to specify the distributions of the
parameters $A$ and $\gamma$. The simplest and most natural assumption
is that these two quantities are not correlated. Consequently, the
distribution  
${\cal P}(A,\gamma)$ can be decoupled as ${\cal P}_A(A)\,{\cal
P}_\gamma(\gamma)$. To specify the distribution ${\cal P}_A$ let us
assume that the EFs are uniformly distributed in space. An EF
behaves like a dipole, either electric or elastic, this allows us to
specify it`s interaction strength as $A(r)=A_0/r^3$, where $r$ is
the distance 
between the contact and a given EF,~\cite{blackhalperin} while $A_0$
is a coupling constant dependent on a specific interaction
mechanism. Note that the quantity $A_0$ has dimension of volume. 
Within this model we arrive at the normalized distribution
function
${\cal P}_A(A)=4\pi A_0/3 {\cal V} A^2$ 
(see Appendix~B for details).      
The distribution ${\cal P}_\gamma(\gamma)$ is specified in a manner which
is commonly used in glasses. Namely, the {\em logarithm} of
$\gamma$ is assumed to be uniformly distributed. Hence, ${\cal
P}_\gamma(\gamma) \propto \gamma^{-1}$, see Appendix~\ref{A}. To
normalize it let us take 
into account that for a given energy spacing $U$ there is a 
{\em maximal} switching rate. Since we are interested in the
fluctuators with $U_i \lesssim kT$, we can specify the maximal
switching rate as $\gamma_T$, which is a function of the
temperature. The actual temperature dependence is determined by the
specific interaction mechanism between the EF and its environment. If
the transitions between the EF states are caused by interaction with
phonons, then $\gamma_T \propto T^3$,~\cite{hunklinger} while if the
transitions  are caused by the electrons excitations, then   $\gamma_T
\propto T$.~\cite{black} Therefore , the
normalized distribution can be specified as ${\cal P}_\gamma
(\gamma)=({\cal L} \gamma)^{-1}$, where ${\cal L}=\ln(\gamma_T/\gamma_{\min})
\gg 1$. Here we have introduced the minimal 
switching rate, $ \gamma_{\min}$. To express the decay in a more clear
form let us introduce the dimensionless frequency $\nu_d$ corresponding to the
interaction strength for an EF 
separated from the contact by an average distance to the active
fluctuators, $\bar{r}\equiv (4\pi P_0kT/3)^{-1/3}$, divided by the
Josephson energy $2eV$. 
We can specify $\nu_d$ as
\begin{equation} \label{eq:nu_d}
\nu_d=4\pi P_0 kT A_0 /3 
=A_0 /\bar{r}^3\, .
\end{equation}
 The decay rate ${\cal K} =-\ln W$ is then given by the expression
\be 
{\cal K}= \frac{\nu_d}{{\cal L}}\int_0^\infty \! \frac{\dif
A}{A^2}\int_{\gamma_{\min}}^{\gamma_T}\! \frac{\dif \gamma}{\gamma}
\left[1-W_s(A, \gamma)\right]\, .
\label{eq:ln_W}
\ee
Note that $g_{fs}$ has to be replaced with $A g_{fm}$ in
the expressions for $W_{s}$ in the
many EF case, because of differences in notation.

To estimate the amount of dephasing let us take into account that
the asymptotic expressions for $W_s$, found in (\ref{eq:W_slow}) and
(\ref{eq:W_fast}) for slow and fast switching respectively, match at
\be
\gamma \tilde{\upsilon}_\zeta \approx (A \bar{g}_\zeta)^2\eta_\zeta /2 \gamma
\, , \label{eq:Sc}
\ee 
where $\tilde{\upsilon}_\zeta$ differs from $\upsilon_\zeta$ defined
in Eq.~(\ref{eq:Uc}) by the replacement $g_{fs} \to Ag_{fm}$ in the
differential equation (\ref{de1}), while
\be
 \bar{g}_\zeta=\frac{1}{2\zeta}\int_{\phi_A}^{\phi_B}
\! g_{fm}(\phi) \, , \
\eta_\zeta \equiv \frac{1}{2\zeta \bar{g}_\zeta^2}
\int_{\phi_A}^{\phi_B}g_{fm}^2 (\phi)\,  \dif \phi\, .
\label{eq:Sc1}
\ee
Note that $\tilde{\upsilon}_\zeta$ is a function of $A$,
$\tilde{\upsilon}_\zeta (A)$. Consequently,
Eq.~(\ref{eq:Sc}) should be treated as an {\emph{equation}} to determine
the characteristic value of the coupling $A$.
Defining the solution of Eq.~(\ref{eq:Sc}) as $A_\zeta$ 
and splitting the integration over $A$ in Eq.~(\ref{eq:ln_W}) as
\be
\int_{\gamma_{\min}}^{\gamma_T}\! \frac{\dif
\gamma}{\gamma}\left(\int_0^{A_\zeta} + \int_{A_\zeta}^\infty
\right)\frac{\dif A}{A^2} \left[1-W_s(A, \gamma) \right]
\label{eq:inta}
\ee
we can use the expression (\ref{eq:W_fast}) in the first interval and
the expression (\ref{eq:W_slow}) in the second one. Since both
integrals are determined by $A_\zeta$ we arrive at the
estimate
\be
- \ln W \approx 3\nu_d \zeta f_\zeta \, , \label{eq:W_many_general}
\ee
where $f_\zeta \equiv  \bar{g}_\zeta \sqrt{\eta_\zeta
\tilde{\upsilon}_\zeta (A_\zeta)}$ is some function of $\zeta$, rather
smooth if $D$ is not close to 1.
We see that the interference pattern decays exponentially with an increasing
distance, $2\zeta$, between the resonances. Generally the 
$\zeta$-dependence of $f_\zeta$  can be
calculated numerically for 
a given transparency $D$ using the analytical expressions (\ref{eq:Uc})   
and (\ref{eq:Sc}).
For a constant $g_{fm}$, $f_\zeta = g_{fm}$. We do not analyze here the
function $f_\zeta$ in detail.

\subsection{Non-optimal EFs}

In the previous sections it has been assumed that the system size
is infinite. A consequence of this assumption is that, independent of
temperature, the EFs which have the strongest effect on the junction
transparency will always be included in the estimates. From the method
used to obtain the general estimate for many EFs,
Eq.~(\ref{eq:W_many_general}), one can conclude that the EFs which
have the most effective coupling fulfill  $A \approx A_\zeta$, this
corresponds to a spatial distance  
$
r_\gamma= (A_0/A_\zeta)^{1/3} \approx
(A_0\bar{g}_\zeta/\gamma)^{1/3}
$.
A further point is that the rate $\gamma$ is confined to the interval
between $\gamma_{\min}$ and $\gamma_T$. Thus we have actually assumed
that the size of the region where EFs reside has a size greater than 
$r_{\max} \approx (A_0\bar{g}_\zeta/\gamma_{\min})^{1/3}$, and that
there is no ``excluded 
region'' without EFs  near the contact with the size less that $r_{\min}
\approx (A_0\bar{g}_\zeta/\gamma_T)^{1/3}$.

What happens if this ``optimum'' EF is out of the  range? This can
occur if the 
system is limited in size, or if there is a specifically pure region
around the contact. 

\subsubsection{Role of finite size of the sample}
Let us first discuss the role of a finite size, $R$ , of the region
containing EFs. 
If $R\lesssim r_{min}=(A_0/A_\zeta)^{1/3}$ then all EFs will act as
``slow'' ones. To estimate $\ln W$ in this case one 
can use the expression $W_s=\cos \Phi_\zeta$ for the whole integration
region over $A$ in Eq.~(\ref{eq:ln_W}), arriving at
\begin{equation} \label{eq:many_slow}
- \ln W = 2 \nu_d \bar{g}_\zeta\zeta \,  F(2A_0\bar{g}_\zeta \zeta/
R^3)    \, . 
\end{equation}
Here 
\be
F(z)=\int_z^\infty \frac{1-\cos x}{x^2} \dif x \, ,
\label{Fz}
\ee
which is a decreasing function of its argument.
This expression (\ref{eq:many_slow}) is valid if its right-hand side
is less the right-hand side of Eq.~(\ref{eq:W_many_general}). 

At intermediate values of $R$, 
 $$(A_0\bar{g}_\zeta/\gamma_{min})^{1/3} \gg R \gg
(A_0\bar{g}_\zeta/\gamma_T)^{1/3}\, , $$ 
as in the previous case, 
only the second integral in the expression (\ref{eq:inta}) does exist,
lowest limit should also be replaced by $A_0
\bar{g}_\zeta/R^3$. However, the approximation $W_s=\cos \Phi_\zeta$
is not valid any more. Using the approximation (\ref{eq:W_slow}) one
can obtain
\begin{equation}
W=\exp\left[-\nu_d \zeta f_\zeta \frac{\ln(\gamma_T R^3/A_0 \bar{g}_\zeta)}{\ln ( \gamma_T/\gamma_{\min})} \right]\, . 
\end{equation}
This means that when the size of system is limited the amount of dephasing can
be less than estimated for an infinite system.

\subsubsection{Role of the spacer}

Let us now discuss the role of a ``pure'' region (spacer) near the
SQPC where there are no EFs. If a typical diameter $r_0$ of such region
is large enough, such that $r_0\gtrsim
r_{\max}=(A_0\bar{g}_\zeta/\gamma_{\min})^{1/3}$, 
then all EFs act as ``fast'' ones.

The single EF solution, $W_{s}$, in this limit
is given by Eq.~(\ref{eq:W_fast}) in Sect.~\ref{single_fast}. 
After calculating the average in Eq.~(\ref{eq:W_h}) we arrive
at
\be
- \ln W \approx (\nu_d \bar{g}_\zeta/{\cal L}) \sqrt{4\pi\zeta\eta_\zeta/\gamma_{\min}}\, .
\label{eq:W_many_fast}
\ee
It is difficult to estimate the actual amount of
dephasing. We have to restrict our conclusions to interpreting how the
amount of dephasing will change depending on the parameters
$\gamma_T$, $\gamma_{\min}$, $\zeta$ and $T$. We already know that
dephasing will increase with temperature. Generally one can also
say that dephasing will increase with $\zeta$. However, at large
enough temperatures when $\gamma_{\min}$ appears large enough,  
the dephasing will decrease. 
The free parameter of the theory, $A_0$, can be estimated only roughly
through comparison with the noise measurements in the normal state. To map
the parameter $A_0$ to the noise let us employ the theory of flicker
noise in a QPC\cite{hg} to the case of a single mode
contact. According to that theory, results for the noise intensity
$S(\tau)$ are substantially dependent on the relationship between the
maximal and minimal distances between the EFs and the QPC. The
simplest case, which is quite realistic, is when these distances are
of the same order of magnitude. When $\Gamma_{T}^{-1} \ll
|\tau| \ll \Gamma_{\min}^{-1}$ the noise intensity can be expressed
as, cf. with Ref. \onlinecite{hg},   
\be
S(\tau)\approx \left({2e^2V \over h}\right)^2 \, \left({4\pi P_0 kT
A_0\over 3}\right)^2 \, \frac{\ln (1/\Gamma_{\min}|\tau|)}{\ln
(\Gamma_{T}/\Gamma_{\min})}\, .
\ee
By obtaining estimates for $\Gamma_{T/\min}$ from noise spectra in
the normal state one can, in principle,  estimate the coupling
parameter $A_0$. A key point is to make measurements of both the
MAQI interference pattern and the normal-state noise spectra in a rather
large frequency range. This combination does not look too simple.

\section{Conclusions}

We have presented a method for investigating the influence of flicker
noise in the junction transparency of a SQPC on coherent Andreev
states. This is done by estimating the effect of these fluctuations on
the so-called microwave-activated quantum interferometer (MAQI)~\cite{prl}.

For a small contact when only a single EF is in range to affect the
junction transparency there can be either a distortion or a decay
of the MAQI interference pattern. A distortion appears for very slow
switching rates of the EF and a weak decay (dephasing) for fast rates. 
It is possible to confirm our model in the fast switching 
limit experimentally.
The only unknown parameter is the switching rate $\gamma$ which 
can be found by measuring the telegraph noise
of the contact in the normal state. This is best done by driving the
system into the normal state with a magnetic field, since $\gamma$ is
temperature dependent.
When $\gamma$ is known it is then possible to calculate the amount of
dephasing and compare with experimental results.

In the limit when the influence of many EFs has to be considered we
have arrived at more general results. It is not possible to make any
exact predictions since the distribution and coupling strength of the
EFs are sample dependent. However, our calculations show that in the 
presence of many
EFs there will always be an exponential decay of the MAQI interference
pattern. The strength of this dephasing should be about the same for
all switching rates. One exception is when
there is an impurity-free region near the SQPC, in this case dephasing will
decrease for higher rates. 

Finally, we note that this paper together with work in
Ref.~\onlinecite{lt22}   
presents a framework which can be used to investigate
the coupling of a SQPC to its electromagnetic environment.

\section*{Acknowledgements}
We are grateful to M. Jonson for fruitful discussions. Support from
the SSF program ``Quantum Devices and
Nanostructures'' and CAS, the Centre for Advanced Study in Oslo,
Norway is acknowledged. 
\appendix

\section{Derivation of equation for $W$}\label{Diff}
In this appendix the differential equation, Eq.~\ref{diff_eq}, for
$W$ is derived. The following is valid for the random telegraph
process $\xi(t)$,
\be
\langle\xi(\phi_1)\xi(\phi_2)\ldots\xi(\phi_n)\rangle=e^{-2\gamma|\phi_2-\phi_1|}\langle\xi(\phi_3)\xi(\phi_4)\ldots\xi(\phi_n)\rangle,
\label{eq:M1}
\ee
when $\phi_1\ge\phi_2\ge\ldots \ge \phi_n$. Further, all averages with
an odd number of $\xi$:s are equal to zero. This follows from the
assumption of equal probability for both EF states.

To facilitate the calculation of $W$, the auxiliary function
\be
\Psi(\phi)=\left< \exp\left[i\int_{\phi_A}^\phi \dif \phi' \, 
g_{fs}(\phi')\, \xi(\phi')\right]\right>\,
\ee 
is defined. By expanding the auxiliary function in a Taylor series and
utilizing the relation in Eq.~\ref{eq:M1} above, it is possible to
rewrite the auxiliary function as
\bea
&&\Psi(\phi)=\nonumber\\
&&1+i^2\int_{\phi_A}^\phi\dif\phi_1\int_{\phi_A}^{\phi_1}\dif\phi_2
g_{fs}(\phi_1)g_{fs}(\phi_2)e^{-2\gamma|\phi_2-\phi_1|}\Psi(\phi_2).
\eea
A derivation with respect to $\phi$ leads to
the following integro-differential equation 
\be
\frac{\dif \Psi}{\dif \phi}=-g_{fs}(\phi)\int_{\phi_A}^\phi\dif \phi'
g_{fs}(\phi')e^{-2\gamma|\phi'-\phi|}\Psi(\phi'),
\label{eq:int_diff}
\ee
and a second derivation provides the following differential equation
\be
\frac{\dif^2\Psi}{\dif \phi^2}+\left[ 2\gamma- \frac{\dif\ln
    g_{fs}(\phi)} {\dif \phi}\right]\frac{\dif \Psi}{\dif \phi} +
g_{fs}^2(\phi)\Psi=0,
\ee
with the initial conditions
\be
\Psi(\phi_A)=1\, , \quad \dif\Psi/\dif\phi|_{\phi=\phi_A}=0\, . 
\ee 
This equation makes it possible to find approximate
analytical solutions for $W$ in the single EF case.

\section{Distributions, ${\cal P}_\gamma(\gamma)$ and ${\cal
P}_A(A)$}\label{A} 

Here we outline the derivation for the distribution functions
${\cal P}(\gamma)$ and ${\cal P}(A)$ which are necessary when averaging in the many
EF case. Let us begin with ${\cal P}(A)$, where $A$ is the coupling strength
between an EF and the contact. By assuming that an EF can be modeled
with a dipole field, we can state that $A\propto A_0/r^3$ where $r$ is
the distance from the EF.\cite{blackhalperin} The normalized distribution is
then found through the 
following integral,
$
{\cal P}(A)={\cal V}^{-1}\int\dif{\bf r}\, \delta(A-A_0/r^3)
$
as 
\be
P(A)=(4\pi/3{\cal V})\, (A_0/A^2)\, .
\label{eq:PA}
\ee 
To find the distribution of $\gamma$ a closer look at the structure of
the two level system (TLS) which is assumed for the EFs is
necessary. The Hamiltonian 
for a TLS can be $H=\Delta \sigma_z-\Lambda\sigma_x$, where
$\Delta$ is the energy difference between the minima of the two
states and $\Lambda$ is the tunneling coupling between the two
states. After diagonalization the excitation energy is found to be
$U=\sqrt{\Delta^2+\Lambda^2}$. By assuming that the tunneling coupling
decays exponentially we have that ${\cal P}(\Lambda)\propto
1/\Lambda$. Further, the hopping rate $\gamma$ depends on
$(\Lambda/U)^2$ and a term $U^3/\hbar E_c^2$ where the latter term
comes from assuming that hopping is phonon mediated ($E_c$ is the
parameter characterizing the 
coupling energy between the phonons and the EFs). Within these 
assumptions ${\cal P}(\gamma)$ is given by,
\bea
{\cal P}_\gamma(\gamma)&=&\int\dif{\Delta}\, \dif\Lambda\,
P(\Lambda)\, \delta(U^2-\Delta^2
-\Lambda^2)\nonumber \\
&& \times
\delta\left(\gamma-\frac{\Lambda^2}{U^2}
\frac{U^3}{\hbar E_c^2}\right)\propto \frac{1}{\gamma}. 
\eea
After normalization we have,
\be
P(\gamma)=\frac{1}{{\cal L} \gamma},
\ee
where ${\cal L}=\ln(\gamma_T/\gamma_{\min})$ with $\gamma_{T,\min}$ being
limiting values for the EF switching rate $\gamma$.

In the general case ${\cal P}={\cal P}(U,\gamma,A)$ and the number of EFs will be given
by $N=\int\dif U\, \dif\gamma\, \dif A\,  {\cal P}(U,\gamma,A)$. By assuming a
constant distribution 
for $U$, which we label ${\cal P}_0$, we arrive at the number of EFs as
$N=P_0{\cal V}kT$. 

The final distribution function for $A$ and $\gamma$  is then
\be
{\cal P}(A,\gamma)=\frac{4\pi P_0 kT}{3}\, \frac{A_0}{A^2}\, 
\frac{1}{{\cal L}\gamma}.
\ee

\end{document}